\documentstyle[12pt,aaspp4,epsf]{article}

\begin{document}

\title{Tidal Decay of Close Planetary Orbits}
\author{F. A. Rasio\altaffilmark{1}}
\affil{Department of Physics, MIT 6-201, Cambridge, MA 02139}
\and
\author {C. A. Tout\altaffilmark{2}, S. H. Lubow and M.\ Livio}
\affil{Space Telescope Science Institute, 3700 San Martin Drive,
Baltimore, MD 21218}
\altaffiltext{1}{Alfred P.\ Sloan Foundation Fellow.}
\altaffiltext{2}{On leave from University of Cambridge, England.}
\authoremail{rasio@mit.edu}

\begin{abstract}
The 4.2-day orbit of the newly discovered planet around 51~Pegasi
is formally unstable to tidal dissipation. 
However, the orbital decay time in this system is longer than the
main-sequence lifetime of the central star.  
Given our best current understanding of tidal interactions, a planet of 
Jupiter's mass around a solar-like star could have dynamically
survived in an orbit with a period as short as $\sim10\,$hr.  
Since
radial velocities increase with decreasing period, we would expect to
find those planets close to the tidal limit first and, unless this is a very
unusual system, we would expect to find many more.  
We also consider the tidal stability of planets around more evolved stars
and we re-examine in particular the question of whether the 
Earth can dynamically survive the 
red-giant phase in the evolution of the Sun.
\end{abstract}


\section{Introduction}

A new era in astronomy has begun recently with the first clear detections 
of several extra-solar planets, first around a millisecond pulsar
(Wolszczan 1994) then around several solar-type stars (Mayor \& Queloz 1995;
Marcy \& Butler 1996). These discoveries will no doubt lead to significant
improvements in our understanding of many processes related not only to planet 
formation, structure and evolution, but also, as we illustrate in this paper,
to stellar astrophysics.

The new planets have brought many surprises. 
In particular, the existence of a Jupiter-type planet with a very short orbital
period of $4.2\,$d around 51~Peg\markcite{ref1} 
(Mayor \& Queloz 1995) is very puzzling. Not only is it difficult
to fit such an object into accepted scenarios for planet formation but
it turns out, as we show below, that its orbit is unstable to 
tidal dissipation. Orbital decay is therefore inevitable in this system, 
and the Jupiter-mass companion will ultimately spiral into the star. 
However, in this system, we will show that the orbital decay timescale 
is longer than the main-sequence lifetime of the solar-like star,
consistent with the planet's surviving to the present.  
We will then investigate tidal
survival in general for Jupiter-like planets in close orbits around 
solar-type stars.  

In the 51~Peg system, we can be sure
that the planet will not survive any post-main-sequence evolution of
the central star.  
By the time the star has grown to about twice its current radius, 
the orbital decay
rate will have exceeded the evolution rate.  This last point brings us
to reconsider the fate of our own planet Earth.  
Current stellar evolution calculations\markcite{ref2} (Sackmann,
Boothroyd, \& Kraemer 1993), taking into account the mass
loss from the Sun and the resulting expansion of the Earth's orbit,
predict that the Earth will not be engulfed even when the Sun reaches its maximum
radius at the tip of the giant branch.
However, as the mass-loss rate increases
with the expansion of the solar envelope, 
so does the tidal decay rate of the Earth's
orbit and we will show here that, as a result, 
the Earth  may well not survive after all.

\section{Tidal Evolution of 51~Peg}

\subsection{Orbital Stability}

The combination of tidal torques and viscous dissipation
in binary systems act both to circularize the orbit and
to synchronize the spin of the stars with the orbital rotation.  
The circular, synchronized state normally corresponds to a
minimum  energy for a given total angular momentum.  However, when
the two components are close, the synchronized state may be
unstable\markcite{ref3} (Hut 1980).
Instability occurs when the ratio of spin angular momentum to orbital
angular momentum $J_{\rm spin}/J_{\rm orb} > 1/3$.

If the 51~Peg system were synchronized at its current orbital period, then,
in terms of observable quantities, we would have
\begin{equation}
{J_{\rm spin}\over J_{\rm orb}} = 1.47\, \left({M\over M_\odot}\right)^{-1/3}
 \left({R\over R_\odot}\right)^2 \left({k\over k_\odot}\right)^2
 \left({P\over 4.23\,{\rm d}}\right)^{-5/3}
 \left({K\over59\,{\rm m}\,{\rm s}^{-1}}\right)^{-1} \sin i, \label{eq1}
\end{equation}
where $M$ and $R$ are the mass and radius of the star, $k=(I_*/MR^2)^{1/2}$
is its dimensionless gyration radius, $P$ is the orbital
period, $K$ is the projected orbital velocity and $i$ is the unknown 
inclination angle of the orbit. The star in 51~Peg appears to be nearly identical 
to our own Sun (mass $M_\odot$, radius $R_\odot$, $k_\odot^2=0.08$; see \markcite{ref1} 
Mayor \& Queloz 1995). From line-profile analysis\markcite{ref4} (Soderblom 1983) and given
that the star is not active\markcite{ref1} (Mayor \& Queloz 1995)
we estimate $\sin i \gtrsim 0.4$. Therefore, if synchronized, 51~Peg is certainly
very close to the formal stability limit and probably exceeds it. 

However, it is almost certain that the star in this system is in fact 
not synchronously rotating. Instead, it appears to be slowly spinning, much as we 
would expect an ordinary solar-like star\markcite{ref1} (Mayor \& Queloz 1995).  
Stars like the Sun are believed to have been spun down by magnetic
braking \markcite{ref5} (Skumanich 1972).  Magnetic dynamo models
indicate that the braking
rate decreases with decreasing angular velocity.  Skumanich's  surface velocity
$v_{\rm surf}\propto t^{-1/2}$, where $t$ is the age of the star, leads
to a spin-down timescale of $\tau_{\rm sd} \propto \Omega_{\rm spin}^{-2}$, while
the theory of \markcite{refTout} Tout \& Pringle (1992) for fully
convective, rapidly rotating stars leads to the weaker, but still inverse, 
dependence of
$\tau_{\rm sd}\propto \Omega_{\rm spin}^{-1/2}$.  In general the tidal
synchronization rate increases with the lack of synchronicity,
$\tau_{\rm su}\propto \Delta\Omega^{-1}$, where $\Delta\Omega$ is the
difference between stellar spin and orbital angular velocities.  If
$\tau_{\rm su}$ and $\tau_{\rm sd}$ were both short compared with the
age of the star then an equilibrium would be set up in which
$\tau_{\rm su} \approx \tau_{\rm sd}$.  This is the case for
cataclysmic variables, in which tidal spin up and magnetic braking (or
gravitational radiation of angular momentum) are in equilibrium
\markcite{refVerbunt} (Verbunt \& Zwaan 1981).  These systems are
unstable because their orbits decay on a timescale $\tau_{\rm
sd}$.  

In the case of 51~Peg, because the star appears to be spinning
slowly we can conclude that $\tau_{\rm sd} \ll \tau_{\rm su}$ over the
lifetime of the star and magnetic braking in this system has proceeded
in a similar way to the Sun.  For the Sun the current braking
timescale is of the order of its age $\tau_{\rm sd}\approx
10^{10}\,$yr so that $\tau_{\rm su}\gg 10^{10}\,$yr.  Further, because
tides transfer orbital angular momentum to the stellar spin,
$\tau_{\rm su}\approx (I_{\rm orb}/I_*) \tau_{\rm a}$, where $\tau_{\rm a}$
is the orbital decay timescale for the system and $I_{\rm orb}$ and
$I_*$ are the moments of inertia the orbit and star respectively, we
may write
\begin{equation}
\tau_{\rm su} = {ma^2\over k^2MR^2}\tau_{\rm a}\approx \tau_a,
\end{equation}
where $m\approx 10^{-3}M$ is the mass of the planet and $a\approx 10R$
is the radius of its orbit.  Thus we expect $\tau_{\rm a}\gg 10^{10}\,$yr
so that, although it is unstable, this particular orbit will not decay
during the star's main-sequence lifetime.  We shall proceed to
calculate the timescale $\tau_{\rm a}$ from detailed tidal theory both
to test this expectation and to place limits on what other systems
might be found in the future.

\subsection{Orbital Decay Rate}

The rate of tidal transfer of angular momentum is
rather uncertain, particularly in solar-like stars where the dissipation
is provided by the eddy viscosity in a relatively
deep but not very massive convective envelope.  However, these
uncertainties do not affect our main conclusions. Assuming standard
tidal dissipation theory (see, e.g., \markcite{ref6}\markcite{ref7} Zahn 1977; Verbunt \&
Phinney 1995) we write the orbital decay rate as
\begin{equation}
{1\over \tau_{\rm a}} = {|\dot a|\over a} =
{f \over \tau_{\rm c}} {M_{\rm env}\over M}\,
 q(1+q) \left({R\over a}\right)^8, \label{eq2}
\end{equation}
where $\tau_{\rm c} \approx (MR^2/L)^{1/3}$ is the eddy turnover timescale. 
Here we will actually use a more precise estimate for the
turnover time of the largest eddies at the {\em base\/} of the convective zone
(cf.\ \S3),
\begin{equation}
\tau_{\rm c} = \left[M_{\rm env}R_{\rm env}(R - R_{\rm env})\over
 3L\right]^{1\over 3}
\end{equation}
where $M_{\rm env}$ is the mass in the convective layer, $R_{\rm env}$
is the radius at its base, and
$M$, $R$ and $L$ are the mass, radius and luminosity of the star.  The
mass ratio $q = m/M$, where $m$ is the mass of the planet and $a$ is
its orbital radius.  The numerical factor $f$ is obtained by
integrating viscous dissipation of tidal energy throughout the
convective zone.  Theoretically\markcite{ref6} (Zahn 1977) and
observationally\markcite{ref7} (Verbunt \& Phinney 1995) $f \approx
1$ as long as
$\tau_{\rm c}\ll P$, the orbital period.  

However, if the tidal pumping period
$P/2$ is less than $\tau_{\rm c}$ then the largest convective cells
can no longer contribute to viscosity, because the velocity field that
they are damping will have changed direction before they can transfer
momentum.  If convection is modelled as a turbulent cascade, then only
eddies which turn over in a time less than the pumping timescale will
contribute\markcite{ref8} (Goldreich \& Keeley 1977) and the average length $l$ 
and velocity $v_{\rm c}$ of
these cells will both be smaller by the same factor $2\tau_{\rm c}/P$. The 
eddy viscosity $\nu \approx v_{\rm c}l/3$
is then reduced by a factor $(2\tau_{\rm c}/P)^2$.  
We may therefore write in general
\begin{equation}
f = f'\,\min\left[1,\left(P\over 2\tau_{\rm c}\right)^2\right],\label{eq4}
\end{equation}
where now $f' \approx 1$.  In a main-sequence solar-type star $\tau_{\rm c} \approx 20\,$d,
so that $f \approx 0.01 f'$.  The planet around 51~Peg 
has a mass similar to that of Jupiter\markcite{ref1} (Mayor \& Queloz 1995) so that 
$q\approx M_{{\rm J}}/M_\odot \approx 10^{-3}$.  
The period of $4.23\,$d then corresponds to a
separation $a \approx 11\,R_\odot$ and we get
\begin{equation}
\tau_{\rm a} \approx 4\times 10^{13}\,{\rm yr}\,\,{1\over f'}
 \left({\tau_{\rm c}\over20\,{\rm d}}\right)^3
 \left(M_{\rm env}\over0.028\,M_\odot\right)^{-1}
 \left({M\over M_\odot}\right)
 \left(q\over 10^{-3}\right)^{-1}
 \left({a \over 11\,R_\odot}\right)^8
 \left({R\over R_\odot}\right)^{-8}, \label{eq5}
\end{equation}
where we have used the value $M_{\rm env} = 0.028\,M_\odot$ from our own solar
models discussed below (\S3).
The star in 51~Peg may be slightly more evolved, and therefore slightly 
larger than the Sun 
at present, with the timescale accordingly shorter.
Mayor \& Queloz quote a radius $R=1.29\,R_\odot$, which would give
$\tau_{\rm a} = 5\times 10^{12}\,$yr.
However, we must compare this orbital decay time with the main-sequence lifetime
$\tau_{\rm ms}\approx10^{10}\,$yr. Over most of this lifetime the star was smaller,
and had a radius closer to $1\,R_\odot$ if its mass is close to $1\,M_\odot$.
Clearly, equation~(\ref{eq5}) shows that $\tau_{\rm a} \gg \tau_{\rm ms}$, consistent with the fact
that this system has survived to the present. 

We can define a minimum period for survival $P_{\rm min}$ by setting 
$\tau_{\rm a}=\tau_{\rm ms}$ in equation~(\ref{eq5}) and solving for $P$. This gives
\begin{equation}
P_{\rm min} = 8.6\,{\rm hr} \left[ f' \left({\tau_{\rm ms}\over10^{10}\,{\rm yr}}\right)
 \left({M_{\rm env}\over0.028\,M_\odot}\right) \left({m\over M_{\rm J}}\right) \right]^{3/10}
 \left({\tau_{\rm c}\over20\,{\rm d}}\right)^{-9/10} \left({M\over M_\odot}\right)^{-7/5}
 \left({R\over R_\odot}\right)^{12/5}
\end{equation}
We see that for 51~Peg we would need $P<P_{\rm min}\approx 9\,$hr before the orbital
decay rate becomes significant.
At this period a synchronized Jupiter of radius $R_{\rm p}\approx R_{\rm J}
\approx 0.1\,R_\odot$ would just fill its Roche lobe (of radius
$R_{\rm L}\approx0.49\, a\, q^{1/3}$ for $q\ll1$) and would therefore begin
to be tidally disrupted anyway. Even if convective viscosity were not
limited by the turnover timescale (i.e., if $f\approx 1$ in equation~(\ref{eq2})), the minimum period
would become $P_{\rm min}\approx2\,$d and 51~Peg's planet would still be safe.  

\begin{figure}
\epsfxsize 6in
\epsffile{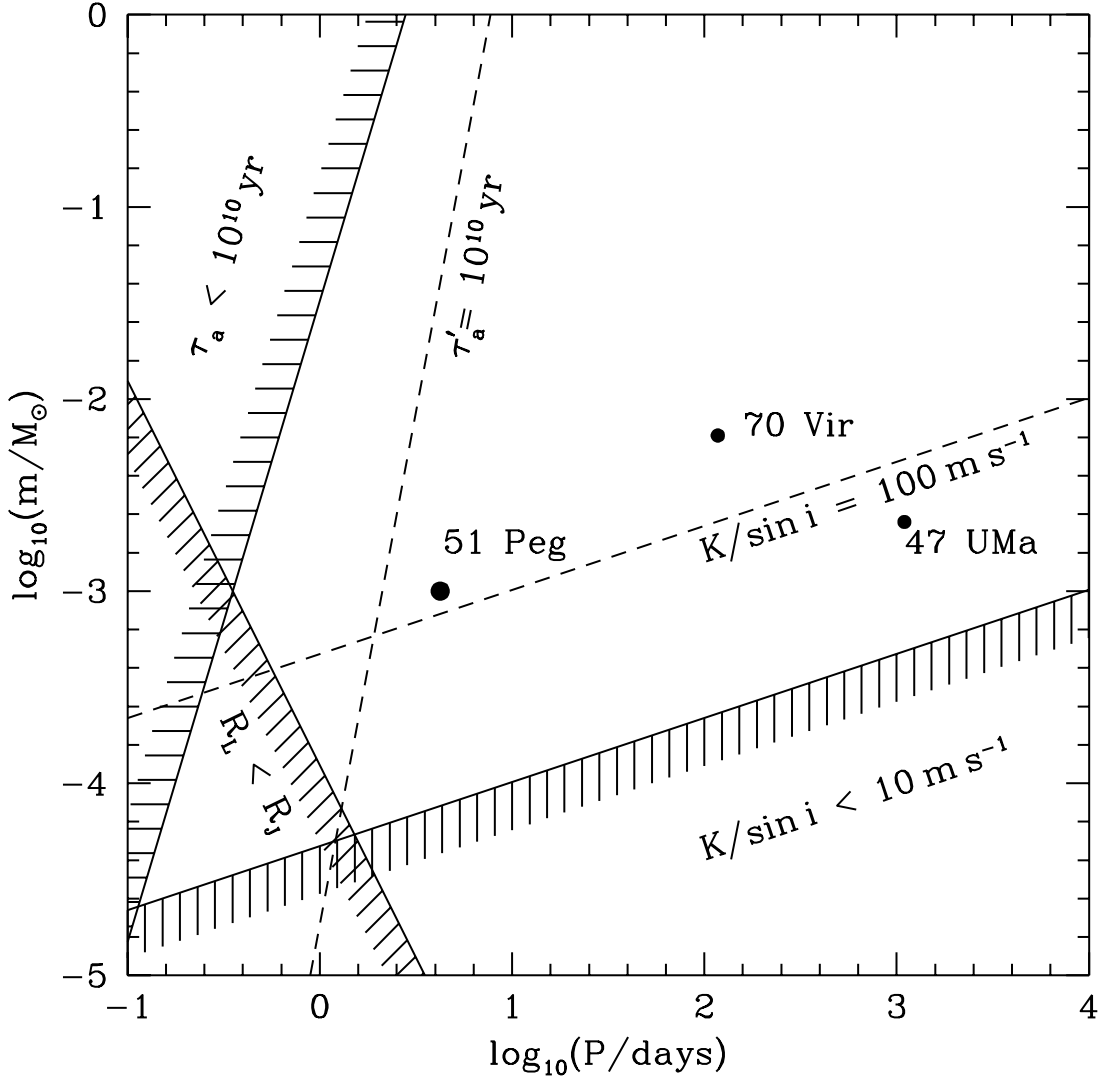}
\caption{The dynamical limits for a planet of mass $m$ and period
$P_{\rm orb}$ around a solar-like star.  The timescale $\tau_{\rm a}$
is the orbital decay timescale obtained from equations~(2)--(5). 
The survival boundary is defined by setting $\tau_{\rm a}=\tau_{\rm ms}$,
where $\tau_{\rm ms}\approx10^{10}\,$yr is the main-sequence lifetime.
The dashed
line for $\tau^\prime_{\rm a}=10^{10}\,$yr applies if the tidal dissipation is not
reduced by the long convective turnover timescale 
(i.e., when setting $f=1$ rather than using eq.~[4]).
Also shown is the Roche limit where
a planet of Jupiter's radius would overflow its critical
tidal lobe ($R_L < R_{\rm J}$).  The boundary where $K/\sin i =
10\,$m$\,$s$^{-1}$ represents an optimistic detection threshold for
current spectroscopic searches. The dashed line at $K/\sin i =
100\,{\rm m\,s}^{-1}$ is a more conservative estimate.
The $M\sin i$ values for the planets around 70~Vir and 47~UMa are also
plotted against their periods.  They are both much further from the
tidal boundaries.}
\end{figure}

In Figure~1 we show the survival boundary for planets around a solar-like
star, as a function of period and planetary mass.
It follows from the figure that we cannot significantly constrain the
mass of this planet by tidal effects.
If planets exist close to this boundary
then they are likely to be found first in radial velocity searches,
because the radial velocity increases with decreasing period.  In the same
figure, we show lines of constant radial velocity, above which
detections are likely.  We note that, although this particular planet
is in the corner of detectability and survivability, it is not so close
to either boundary as to make it unlikely.
The positions of the more recently discovered 
planets around 70~Vir and 47~UMa (Marcy \& Butler 1996)
are also included in the figure.  Both are further from
the tidal boundaries than 51~Peg.

\subsection{Orbital Circularization}

The current measurement of the orbital eccentricity of 51 Peg gives
$e=0.09\pm0.06$, indicating a nearly-circular orbit (Mayor \& Queloz 1995).
We can be fairly certain that
tidal dissipation {\em in the star\/} has not 
played any role in the circularization
of the orbit during the main-sequence phase of the evolution. 
This is because the timescales for circularization 
and orbital decay are the same within a 
factor of 2 (see, e.g., Zahn 1977). Therefore,
the same mechanism would have brought
about a substantial decay of the orbit, which would now be shrinking
rapidly, making it highly improbable that the system would be
found in its current state.

However, the star also raises tides on the planet.  These tides can synchronize
the planet's spin with the orbit in a timescale\markcite{ref12}
(Guillot et al.\ 1995)
\begin{equation}
\tau_{\rm s}\approx Q\left(R_{\rm p}^3\over
Gm\right)\omega\left(m\over M\right)^2\left(a\over R_{\rm
p}\right)^6,
\end{equation}
where $\omega = |\omega_{\rm spin} - \omega_{\rm orb}|$ is the
difference between the spin and orbital angular velocities.  The factor $Q$ is
inversely proportional to the dissipation and has been determined
observationally\markcite{ref13} (Ioannou \& Lindzen 1993) for Jupiter to be about $10^5$.  Therefore the
timescale to synchronize a Jupiter-like
planet originally spinning as fast as Jupiter is $\tau_{\rm s} \approx
2\times 10^6\,$yr. The same tides can
also circularize the orbit, even after synchronization, because radial
tides can dissipate energy without transferring angular momentum.  The
circularization timescale is\markcite{ref14} (Goldreich \& Soter 1966)
\begin{equation}
\tau_{\rm e} = {4\over 63}Q\left(a^3\over GM\right)^{1\over
2}\left(m\over M\right)\left(a\over R_{\rm p}\right)^5.
\end{equation}
Now $Q$ is proportional to the tidal pumping period, so that, with the
same dissipation mechanism, it will be ten times larger after
synchronization of the planet.  The circularization
timescale becomes $\tau_{\rm e}\approx 2\times
10^9\,$yr.  Since this is less than the age of the system, the fact
that the observed orbit is circular is not surprising.

\section{Implications for Planet Formation}

Clearly, if planets
commonly form at short periods, and we assume that they can survive
the effects of thermal evaporation (Guillot et al.\ 1996), 
we may now expect to find many more.
Such discoveries can affect our ideas of planet
formation \markcite{ref9} (see, e.g., Lissauer 1993 and references therein),
which require Jovian
planets to form at distances of a few astronomical units or
more.

Indeed, models of the proto-solar nebula suggest that the temperature
increases very rapidly close to the center. This rapid rise in
temperature has been invoked to explain the difference in composition
between rocky terrestrial planets and ice-rich satellites in the outer solar
system, as well as the absence of objects inside Mercury's orbit.
The high temperature is thought to prevent the condensation of high-Z
material into grains \markcite{refBarshay}
(Barshay \& Lewis 1976), thereby preventing planetesimal
formation.
In addition, any planetesimals that would form in the innermost region of a
protoplanetary  disk are expected to spiral-in rapidly
because of gas drag, and to be destroyed.
Alternatively, models in which the formation of a giant
planet occurs directly by gravitational collapse in a massive disk
have also been discussed, but recently these models have fallen out of
favor for a variety of reasons (Lissauer 1993).

If the new planet in 51~Peg had formed, like our own Jupiter, at a large
distance from the central star, some
angular-momentum-loss mechanism must have brought it in.  Any
{\em dissipative\/} mechanism, such as friction in a primordial stellar nebula,
would, like the tides, increase rapidly with decreasing separation.
It would have had to switch off at a critical moment
for the planet to end up so close to the star and, in this case, this
kind of system ought to be extremely rare. 

An interesting formation process based on a dissipative mechanism has been
discussed by Lin, Bodenheimer \& Richardson (1996).
The mechanism they consider is the dissipative interaction between the
planet and a protoplanetary disk. Spiraling-in of the planet could be halted 
at a small distance from the star when the tidal interaction becomes
important. For this to work, Lin et al.\ must invoke a somewhat
shorter timescale for tidal interaction on the pre-main-sequence. 
In addition, the star must be spinning faster
than the planet is orbiting while the external torque acts.  
We note that, since the star is spinning much more slowly today, its spin 
rate must have passed through a point
of synchronous rotation and become unstable some time in the past. The system
could only have survived this if the star had already contracted or altered its
structure sufficiently that $\tau_a$ were already long at that time.  This
would require a rather delicate balance between the spin-down, contraction and
tidal timescales of the protostellar 51~Peg.
For instance, a spin-down timescale of the order of $10^4\,$yr,
deduced from the theory of Tout \& Pringle (1992), would have left the
structure and size of the protostar virtually unchanged at corotation and the
planet would inevitably have fallen in.

Alternatively, a {\em dynamical\/}
mechanism for angular momentum loss could also be invoked. 
One possibility (Rasio \& Ford 1996)
is that two (or more) Jupiter-like planets 
had initially formed (in the conventional way) at a large distance 
from the central star, and later interacted dynamically.
This could happen because the planets' orbits evolved secularly at different 
rates, or because their masses increased, resulting in
a dynamical instability of the orbits and a close interaction between
the two planets (Gladman 1993).
The interaction can lead to the ejection of one planet, leaving the
other in a highly eccentric orbit. If the pericenter distance of the
inner planet is sufficiently small, its orbit can later circularize at an
orbital separation of a few stellar radii, as seen in 51~Peg. Otherwise,
the planet would be left in a highly eccentric orbit at some intermediate
distance, as seen in another recently discovered system, 70~Vir
(Marcy \& Butler 1996).

\section{Implications for Tidal Dissipation Theory}

If common, planets such as the one around 51~Peg will be invaluable for
improving our understanding of tidal dissipation. Their advantage
over more equal mass binary systems lies in the asynchronicity of the
star leading to orbital decay, rather than just to circularization.  
For instance, we
might envisage a more effective source of viscosity than convective
turbulence, such as magnetic fields generated by the tidal
disturbance.  This would move the tidal boundary in Figure~1, and the
discovery of planets well to the left of the new boundary could be
used to rule out that particular mechanism.  

Even with this system we
can already comment on a much more efficient mechanism for
circularization discussed by Tassoul
and Tassoul\markcite{ref10} (Tassoul \& Tassoul 1992).  In their
model, tidal forces induce circulations
which are viscously damped at the effectively rigid surface of the
star.  The timescale for orbital decay is\markcite{ref11} (Livio 1994)
\begin{equation}
\tau_{\rm a}\approx 9\times 10^6\,{\rm yr}\,(1+q)^{-11/8}\left(L\over
L_\odot\right)^{-{1\over 4}}\left(M\over M_\odot\right)^{-{1\over
8}}\left(M_{\rm env}\over 0.028\,M_\odot\right)^{-1}\left(R\over
R_\odot\right)^{9\over 8}
\left(a\over 11\,R\right)^{49\over 8},
\end{equation}
for 51~Peg.  
Even if a factor of 100 (from eq.~[\ref{eq4}]) is applicable here too, this is
still much less than the age of the system. Therefore we 
can deduce observationally that the Tassoul
mechanism does not operate in this system.

\section{Planets around Red Giants}

\subsection{The Fate of the Earth}

The star 51~Peg is already slightly more evolved than the
Sun\markcite{ref1} (Mayor \& Queloz 1995).  
When it becomes a red giant of about twice the Sun's radius,
it will already possess a
deep, massive convective envelope in which the critical turnover time
is not much longer than it is now.  The orbital decay time $\tau_{\rm a}$ 
will then be reduced by a factor of about $10^4$, becoming shorter than the stellar
evolution time for the red giant. Inevitably, the planet's orbit will then 
decay into the stellar envelope.

This last point brings us to re-examine the fate of our own Earth.
In the case of a planetary system resembling our own, the planets can get close
enough to the central star only when the star evolves to become a giant.
The innermost planets may be completely engulfed inside the stellar envelope
during this
phase. Early stellar evolution calculations suggested that this would in fact
happen to the Earth. First, on
the red giant branch, the Sun would swell up, almost but not quite
reaching the Earth's orbit. Then,
on the asymptotic giant branch, after the exhaustion of core helium, the
Sun would engulf the Earth altogether. 

However, theoretical
considerations\markcite{ref15} (Faulkner 1972) and
observations\markcite{ref16} (Reimers 1975) of giants
and supergiants indicate that these stars suffer mass loss at high rates.
As a result of mass loss, the Earth's orbit expands owing to conservation
of angular momentum. An inclusion of this effect in the evolutionary
calculations for the Sun\markcite{ref2} (Sackmann et al.\ 1993) 
show that its mass will be
reduced to $0.59\,M_\odot$
when it reaches its maximum radius of $0.99\,$AU. By this
time, however, the earth's orbit will be at $1.69\,$AU
and so the current prediction is that the Earth will in fact
survive this phase\markcite{ref17} (see also Maddox 1994).  
A reduction in the mass-loss rate (which is somewhat uncertain)
would work in the direction of increasing the probability of engulfing (by
increasing the Sun's final radius and reducing the expansion of the 
Earth's orbit), but it
appears that for mass-loss rates which are consistent with those
deduced from
star clusters\markcite{ref18} (Weidemann \& Koester 1983), the Earth would always survive.
The important point, however, is that the 
calculations of Sackmann et al.\markcite{ref2} (1993) neglected
the orbital decay driven by the 
tidal interaction between the Earth and the Sun.

\subsection{Effects of Tidal Dissipation}

We have recomputed the evolution of the 
Earth--Sun system, using the evolutionary code of
Eggleton\markcite{ref19} (Pols et al.\ 1995),
taking into account  the effects of tidal dissipation. 
Mass loss was included according to the Reimers
formula\markcite{ref20} (Kudritzki \& Reimers 1978)
\begin{equation}
{\dot M}=-4\eta 10^{-13}\,M_\odot{\rm yr}^{-1} {(L/L_\odot)(R/R_\odot)
 \over (M/M_\odot)},
\end{equation}
with $\eta=0.6$, as in\markcite{ref2} Sackmann et al.\ (1993).
First, we confirm their results: when we neglect the
tidal interaction, we find that the Earth always remains safely outside
the Sun's expanding envelope. At the tip of
the giant branch we find $a = 315\,R_\odot$, while the evolved solar
radius $R^\prime_\odot = 195\,R_\odot$.
The evolution of the radius of the Sun and the radius of the Earth's orbit are
shown in Figure~2. 

Next, we include the tidal interaction by integrating
equation~(\ref{eq2}) in parallel
with the stellar evolution equations. Here we can assume $f\approx1$ since
$P > 1\,{\rm yr} > \tau_{\rm c}$.
For $f=1$ (as favored for example
by the results of Verbunt \& Phinney\footnote{Note, 
however, that our formulation of eq.~\ref{eq2} will
slightly overestimate the tidal effects relative
to the results of\markcite{ref7} Verbunt \& Phinney (1995) with the same 
numerical value for $f$.} 1995), 
we find that the effects of tidal dissipation are already quite
noticeable at the tip of the giant branch.
Note that Venus' orbit would be engulfed by the Sun already in this case
(see Fig.~2). 
For $f=2$ the Earth's orbit decays rapidly enough for our planet 
to be completely engulfed as well.
Our current understanding of tidal dissipation is
sufficiently uncertain that $f$ could well be as large as this, especially
if magnetic fields are important. Clearly, any additional
observational evidence from nearby planets in close orbits could
be invaluable in determining the fate of our own.

\begin{figure}
\epsfxsize 6in
\epsffile{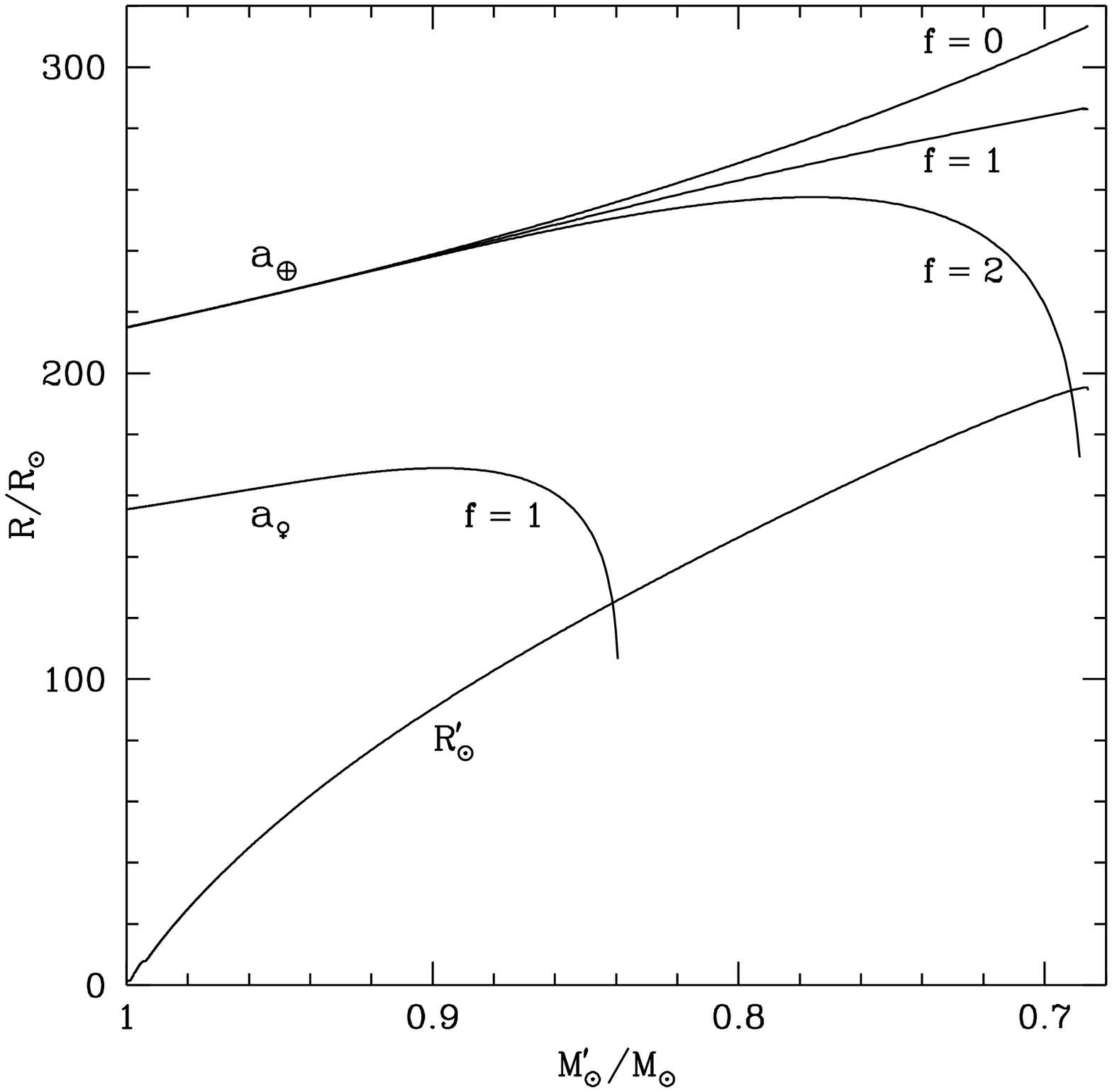}
\caption{The evolution of the orbital separation of the Earth $a_\oplus$ and the
radius of the Sun $R'_\odot$ as functions of its mass $M'_\odot$,
which is falling (increasingly rapidly) as a
result of mass loss in a stellar wind.
The top line shows the Earth's orbit in the absence of tidal dissipation.  
The next two
lines correspond to tidal dissipation with $f=1$ and
$f=2$ in equation~(2).  The orbit of Venus
is also shown with tidal
effects included for $f=1$.  When the orbit curves intersect the solar-radius
curve the Sun's expanding envelope engulfs the planets. 
For $f=2$ this will happen to the Earth in $8\,$Gyr from now.}
\end{figure}

\acknowledgments

We thank Doug Lin for useful discussions.
F.A.R.\ is supported by an Alfred P.\ Sloan Research Fellowship.
S.L.\ acknowledges support from NASA Grant NAGW-4156 and M.L.\ from 
NASA Grant NAGW-2678.

\end{document}